\documentclass[12pt,letterpaper]{article}

\usepackage{natbib}
\usepackage[ left=1in, top=1in, right=1in, bottom=1in]{geometry}
\usepackage{xcolor,graphicx,bm,bbm,colonequals,amsmath,amssymb,url}
\usepackage{array,tabularx,multirow}
\usepackage{enumitem,algpseudocode}
\usepackage[font={footnotesize}]{caption,subcaption}
\usepackage{color}
\usepackage{caption}
\usepackage{subcaption}

\usepackage[normalem]{ulem} 
\usepackage{bbm}

\bibpunct[, ]{(}{)}{;}{a}{,}{,}

\usepackage{mathtools}
\mathtoolsset{showonlyrefs}

\usepackage{amsthm}
\newtheoremstyle{propstyle} 
    {3mm}                    
    {1mm}                    
    {\itshape}                   
    {}                           
    {\scshape}                   
    {.}                          
    {.5em}                       
    {}  
\theoremstyle{propstyle}

\theoremstyle{propstyle}

\theoremstyle{propstyle}

\newsavebox\ideabox

\newcommand{\bs}{\mathbf{s}}

\newcommand{\by}{\mathbf{y}}

\newcommand{\bw}{\mathbf{w}}
\newcommand{\bu}{\mathbf{u}}

\newcommand{\bz}{\mathbf{z}}

\newcommand{\bA}{\mathbf{A}}
\newcommand{\bY}{\mathbf{Y}}

\newcommand{\bG}{\mathbf{G}}

\newcommand{\bI}{\mathbf{I}}
\newcommand{\bD}{\mathbf{D}}
\newcommand{\bH}{\mathbf{H}}
\newcommand{\bU}{\mathbf{U}}
\newcommand{\bV}{\mathbf{V}}

\newcommand{\bX}{\mathbf{X}}

\newcommand{\bfzero}{\mathbf{0}}

\newcommand{\bftheta}{\bm{\theta}}

\newcommand{\bfSigma}{\bm{\Sigma}}

\DeclareMathOperator{\var}{var}

\DeclareMathOperator{\diag}{diag}

\DeclareMathOperator{\KL}{KL}

\DeclareMathOperator{\tr}{tr}

\newcommand{\normal}{\mathcal{N}}
\newcommand{\order}{\mathcal{O}}

\newcommand{\domain}{\mathbb{D}} 

\title{Bayesian nonstationary and nonparametric covariance estimation for large spatial data}

\author{Brian Kidd\thanks{Department of Statistics, Texas A\&M University} \and Matthias Katzfuss\footnotemark[1] \thanks{Corresponding author: \texttt{katzfuss@gmail.com}}}

\date{}

\begin{document}

\maketitle

\begin{abstract}
In spatial statistics, it is often assumed that the spatial field of interest is stationary and its covariance has a simple parametric form, but these assumptions are not appropriate in many applications. Given replicate observations of a Gaussian spatial field, we propose nonstationary and nonparametric Bayesian inference on the spatial dependence. Instead of estimating the quadratic (in the number of spatial locations) entries of the covariance matrix, the idea is to infer a near-linear number of nonzero entries in a sparse Cholesky factor of the precision matrix. Our prior assumptions are motivated by recent results on the exponential decay of the entries of this Cholesky factor for Mat\'ern-type covariances under a specific ordering scheme. Our methods are highly scalable and parallelizable. We conduct numerical comparisons and apply our methodology to climate-model output, enabling statistical emulation of an expensive physical model.
\end{abstract}


\section{Introduction \label{sec:intro}}

\begin{figure}
    \centering
    \includegraphics[width=\textwidth]{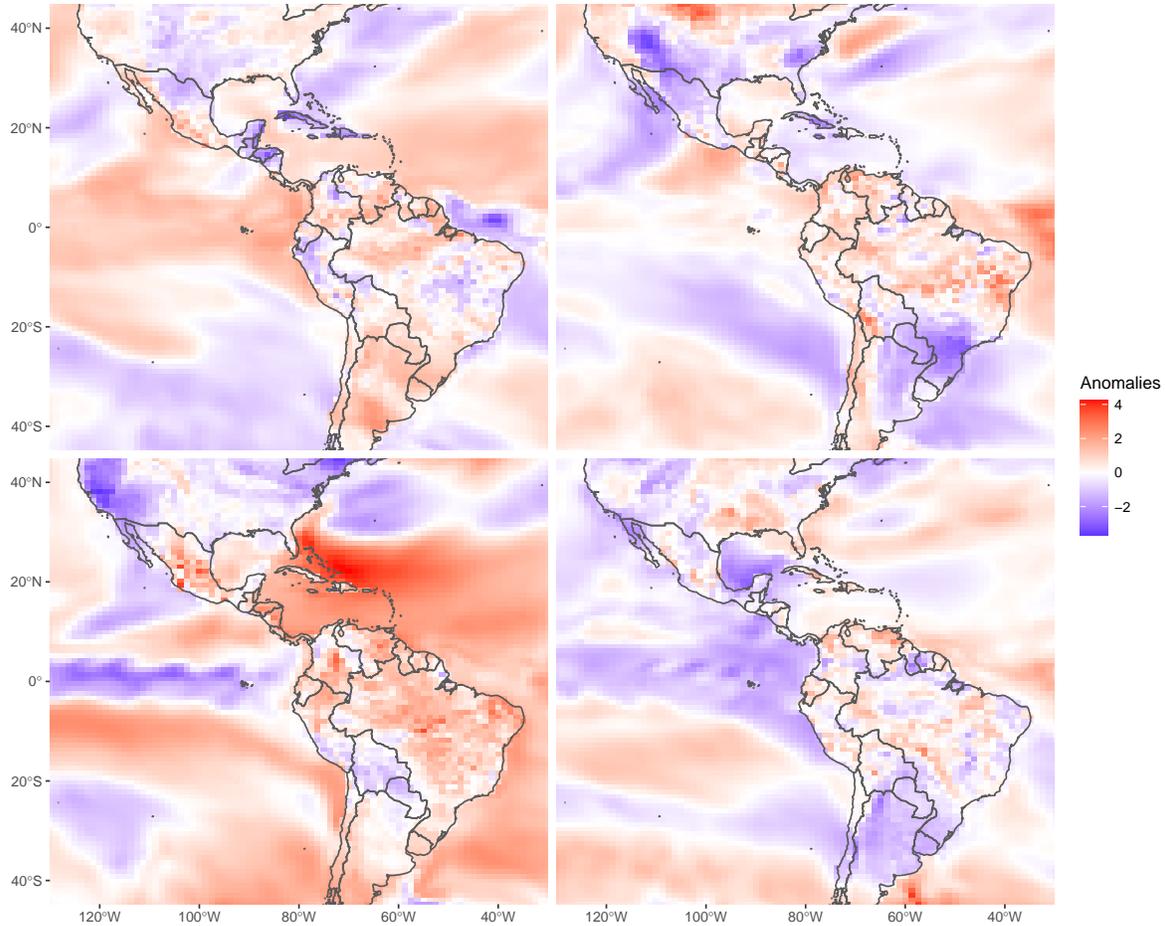}  \caption{Four members of an ensemble of surface-temperature anomalies (in Kelvin) produced by a climate model, on a grid of size $n = 81 \times 96 = 7{,}776$ (see Section \ref{sec:data} for more details)}
\label{fig:tempens}
\end{figure}

Modeling spatial data typically involves specification of spatial dependence in the form of a covariance function or matrix, under an implicit or explicit assumption of joint Gaussianity. 
A motivating example for this paper is statistical climate-model emulation \citep[e.g.,][]{Castruccio2013,Castruccio2014,Nychka2018,Haugen2019}: based on an ensemble of spatial fields generated by an expensive computer model (Figure \ref{fig:tempens}), the goal is to learn the underlying joint  distribution, and then, for instance, to draw additional samples from the distribution. This involves many challenges, including small ensemble sizes, high-dimensional distributions, and complex, nonstationary dependence.
Thus, there is a need for flexible and scalable methods for inferring high-dimensional spatial covariances.

Countless approximations have been proposed to address computational challenges in spatial statistics \citep[see][for a recent review and comparison]{Heaton2017}. In recent years, there has been increasing interest in the idea of \citet{Vecchia1988}, which effectively approximates the Cholesky factor of the precision (i.e., inverse covariance) matrix as sparse. Under certain settings, the Vecchia approximation can provably provide $\epsilon$-accurate approximations at near-linear computational complexity in the number of spatial locations \citep{Schafer2020}. A generalization of the Vecchia approach includes many popular spatial approximations as special cases \citep{Katzfuss2017a}. However, Vecchia approaches have mostly been used for approximating parametric and often isotropic covariance functions.

Isotropic, parametric covariance functions (e.g., Mat\'ern) only depend on spatial distance and on a small number of unknown parameters. Despite being highly restrictive, this is the standard assumption in spatial statistics, especially in the absence of replicates. Approaches to relax these assumptions include parametric nonstationary covariances \citep[e.g., as reviewed by][]{Risser2016}, stationary nonparametric covariances \citep[e.g.,][]{Huang2011,Choi2013,Porcu2019}, nonparametric and nonstationary covariances \citep[e.g.,][]{Fuentes2002}, and domain transformations \citep[e.g.,][]{sampson1992nonparametric,damian2001bayesian,Qadir2019}. In the context of local kriging, covariance functions are typically estimated locally from a parametric \citep[e.g.,][]{Anderes2011,Nychka2018} or nonparametric \citep[e.g.,][]{Hsing2016} perspective, but this generally does not imply a valid joint model or positive-definite covariance matrix.

Outside of spatial statistics, covariance estimation is often performed based on (modified) Cholesky decompositions of the precision matrix. This approach is attractive, because it automatically ensures positive-definiteness, because sparsity in the Cholesky factor directly corresponds to ordered conditional independence and hence to directed acyclic graphs, and because it allows covariance estimation to be reformulated as a series of regressions. Regularization can be achieved as in other regression settings, for example by enforcing sparsity using a Lasso-like penalty or a thresholding procedure \citep[e.g.,][]{huang2006covariance, Levina2008} 
or via Bayesian prior distributions \citep[e.g.,][]{smith2002parsimonious}. 
Motivated by a Gaussian Markov random field assumption for spatial data, \citet{zhu2009estimating} estimate the Cholesky factor based on an ordering of the spatial locations intended to minimize the bandwidth, which amounts to coordinate ordering on a regular grid, and they regularize the entries of the Cholesky factor using a weighted Lasso penalty depending on spatial distance; this approach scales cubically in the number of spatial locations.

Here, we propose scalable nonparametric Bayesian inference on a high-dimensional spatial covariance matrix. 
The basic idea is to infer a near-linear number of nonzero entries in a sparse Cholesky factor of the inverse covariance matrix. Our model can be viewed as a nonparametric extension of the Vecchia approach, as regularized inference on a sparse Cholesky factor of the precision matrix, or as a series of Bayesian linear regression or spatial prediction problems. We specify prior distributions that are motivated by recent results \citep{Schafer2017,Schafer2020} on the exponential decay of the entries of the inverse Cholesky factor for Mat\'ern-type covariances under a maximum-minimum-distance ordering of the spatial locations \citep{Guinness2016a,Schafer2017}. 
Our method scales well to very large datasets, as the number of nonzero entries in the Cholesky factor and the computational cost both scale near-linearly in the number of spatial locations, in effect inferring a near-linear number of parameters in the sparse inverse Cholesky factor instead of a square number of parameters in the dense covariance matrix. Further speed-ups are possible, as the main computational efforts are perfectly parallel. Our approach is applicable to a single realization of the spatial field, but the inference will be most useful and accurate if replicate observations are available. 

The remainder of this document is organized as follows. Section \ref{sec:methodology} describes our methodology. Section \ref{sec:simulation} provides numerical comparisons using simulated data. In Section \ref{sec:data}, our method is used for climate-model emulation. Section \ref{sec:conclusions} concludes.

\section{Methodology \label{sec:methodology}}

\subsection{Sparse inverse Cholesky approximation for spatial data\label{sec:model}}

\begin{figure}
    \centering
    \includegraphics[width=\textwidth]{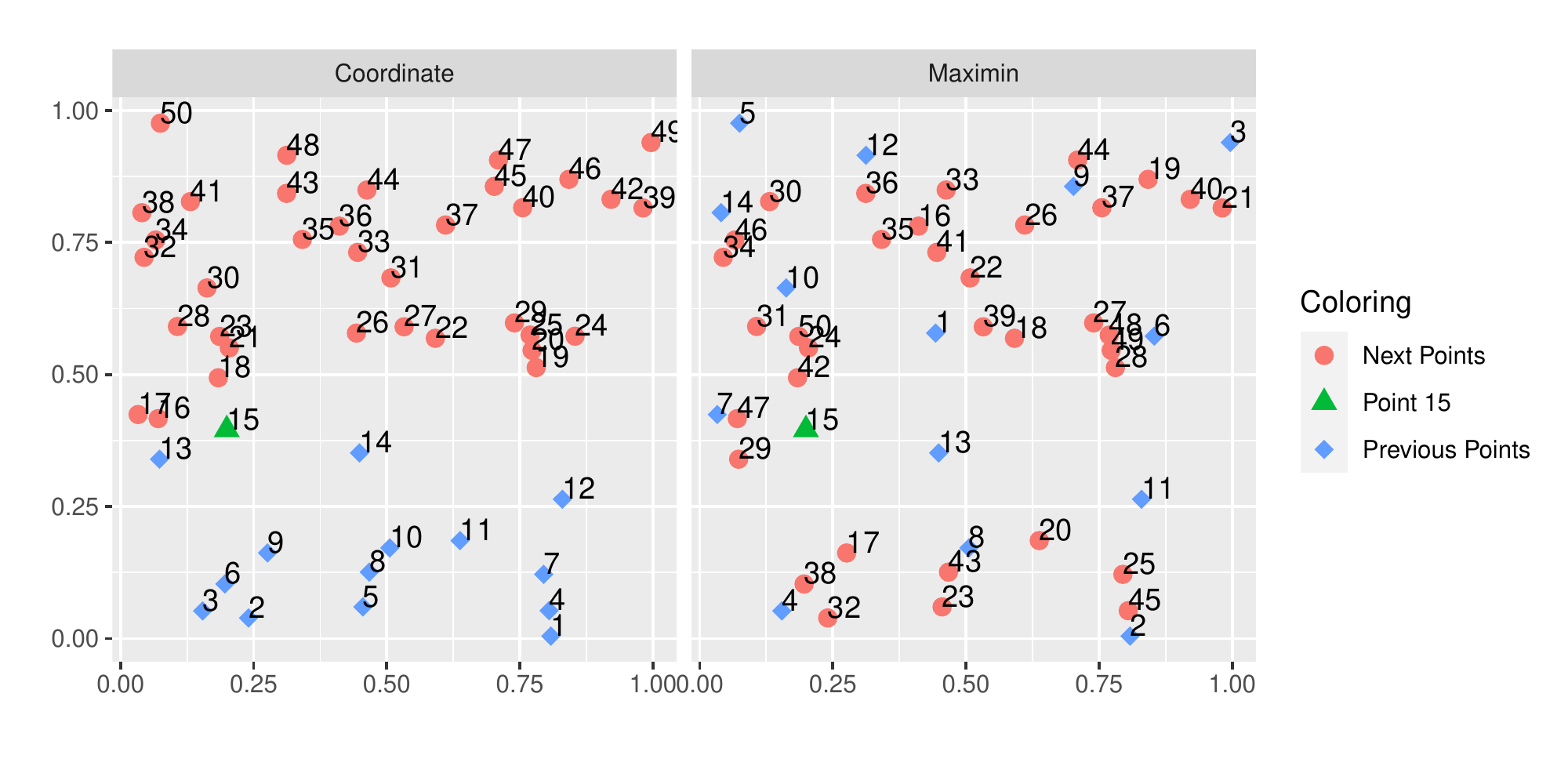}
    \caption{For $n=50$ randomly sampled locations on the unit square, comparison of coordinate (bottom to top) and maximin ordering. For $i=15$, previously ordered locations $\bs_1,\ldots,\bs_{n-1}$ are highlighted in blue to show their roughly equidistant spread over the domain for maximin. As an example, for $m=4$, we would have conditioning sets $g_4(15) = (13,9,14,6)$ for coordinate and $g_4(15) =(7,13,10,1)$ for maximin.}
    \label{fig:ordering}
\end{figure}

Consider a $N \times n$ matrix of spatial data, 
\begin{equation}
    \label{eq:datamat}
\bY = 
\begin{pmatrix} 
y_1^{(1)} & \cdots & y_n^{(1)} \\
\vdots & \ddots & \vdots \\
y_1^{(N)} & \cdots & y_n^{(N)}
\end{pmatrix}
=
\begin{pmatrix} 
\text{---} \!\!\! & \by^{(1)}{}' & \!\!\! \text{---} \\
& \vdots & \\
\text{---} \!\!\! & \by^{(N)}{}' & \!\!\! \text{---}
\end{pmatrix}
=
\begin{pmatrix} 
\vert &  & \vert \\
\by_1 & \cdots & \by_n \\
\vert &  & \vert
\end{pmatrix},
\end{equation}
where $y_i^{(\ell)}$ is the $\ell$th observation at spatial location $\bs_i$. 
We assume that the locations $\bs_1,\ldots,\bs_n$, and hence the columns of $\bY$, are ordered according to a maximin ordering \citep{Guinness2016a,Schafer2017}, which sequentially selects each location in the ordering to maximize the minimum distance from locations already selected (see Figure \ref{fig:ordering}).

We model the rows $\by^{(\ell)} = (y_1^{(\ell)},\ldots,y_n^{(\ell)})'$ of $\bY$ as independent $n$-variate Gaussians:
\begin{equation}
\label{eq:datamodel}
\by^{(\ell)} | \bfSigma \stackrel{iid}{\sim} \normal_n(\bfzero,\bfSigma), \qquad \ell = 1,\ldots,N.
\end{equation}
We assume that the data are centered, either using an ad-hoc pre-processing step (e.g., by subtracting location-wise means) or using a more elaborate procedure (see Section \ref{sec:noise}).

Our goal is to make inference on the $n \times n$ spatial covariance matrix $\bfSigma$ based on the $N \times n$ observations $\bY$, in the case where $n$ is large (at least in the thousands) and $N$ is relatively small. Typically, a parametric, and often isotropic, covariance function is assumed such that $\bfSigma$ is a function of only a small number of parameters, which can then be estimated relatively easily. Here, we avoid explicit assumptions of stationarity and isotropy. 

We assume a form of ordered conditional independence,
\begin{equation}
\label{eq:screening}
p(y_i^{(\ell)}|\by_{1:i-1}^{(\ell)},\bfSigma) = p(y_i^{(\ell)}|\by_{g_m(i)}^{(\ell)},\bfSigma), \qquad i=2,\ldots,n, \quad \ell =1,\ldots,N,
\end{equation}
where $g_m(i) \subset (1,\ldots,i-1)$ is an index vector consisting of the indices of the $\min(m,i-1)$ nearest neighbors to $\bs_i$ among those ordered previously; that is, $\bs_{(g_m(i))_j}$ is the $j$th nearest neighbor of $\bs_i$ among $\bs_1,\ldots,\bs_{i-1}$ (see Figure \ref{fig:ordering}). While \eqref{eq:screening} holds trivially for $m = n-1$, for many covariance structures it even holds (at least approximately) for $m \ll n$, as has been demonstrated numerically \citep[e.g.,][]{Vecchia1988, Stein2004,Datta2016,Guinness2016a,Katzfuss2017a,Katzfuss2018,Katzfuss2020} and theoretically \citep{Schafer2020} in the context of Vecchia approximations of parametric covariance functions. Assume for now that $m$ is known.

Consider the modified Cholesky decomposition of the precision matrix:
\begin{equation}
\label{eq:cholesky}
\bfSigma^{-1} = \bU \bD^{-1}\bU',
\end{equation}
where $\bD = \diag(d_1,\ldots,d_n)$ is a diagonal matrix with positive entries $d_i > 0$, and $\bU$ is an upper triangular matrix with unit diagonal (i.e., $\bU_{ii}=1$). (To be precise, \eqref{eq:cholesky} is the reverse-ordered Cholesky factorization of the reverse-ordered $\bfSigma^{-1}$, which simplifies our notation later.) The ordered conditional independence assumed in \eqref{eq:screening} implies that $\bU$ is sparse, with at most $m$ nonzero off-diagonal elements per column \citep[e.g.,][Prop.~3.1]{Katzfuss2017a}. We define $\bu_i = \bU_{g_m(i),i}$ as the nonzero off-diagonal entries in the $i$th column.

\subsection{Covariance estimation via Bayesian regressions\label{sec:predictive}}

From \eqref{eq:cholesky}, we see that we can estimate the $\order(n^2)$ unknown entries of $\bfSigma$ by inferring the $\order(nm)$ variables $d_1,\ldots,d_n$ and $\bu_1,\ldots,\bu_n$. To do so, our data model \eqref{eq:datamodel} can be written as a series of $n$ linear regression models \citep{huang2006covariance}:
\begin{equation}
\label{eq:regression}
p(\bY | \bfSigma) = \prod_{i=1}^n p(\by_i|\by_{1:i-1} , \bfSigma) = \prod_{i=1}^n \normal_N(\by_i|\bX_i\bu_i,d_i\bI_N),
\end{equation}
where the ``response vector'' $\by_i = (y_i^{(1)},\ldots,y_i^{(N)})'$ is the $i$th column of $\bY$ in \eqref{eq:datamat} consisting of the $N$ observations at the $i$th spatial location, and the ``design matrix'' $\bX_i$ consists of the observations at the $m$ neighbor locations of $\bs_i$, stored in the columns of $\bY$ with indices $g_m(i)$; specifically, $\bX_i$ is an $N \times m$ matrix with $\ell$th row $-\by_{g_m(i)}^{(\ell)}{}'$. 

The Bayesian regression models in \eqref{eq:regression} are completed by independent conjugate normal-inverse-gamma (NIG) priors:
\begin{equation}
    \label{eq:nig}
 \bu_i | d_i,\bftheta \stackrel{ind.}{\sim} \normal(\bfzero,d_i\bV_i), \qquad d_i | \bftheta \stackrel{ind.}{\sim} \mathcal{IG}(\alpha_i,\beta_i), \qquad i=1,\ldots,n,
\end{equation}
where $\bftheta$ is a vector of hyperparameters determining $m$, $\bV_i$, $\alpha_i$, and $\beta_i$ (see Section \ref{sec:shrink} below).
Due to conjugacy, the posterior distributions (conditional on $\bftheta$) are also NIG:
\begin{align}
p(\bu_1,\ldots,\bu_n,d_1,\ldots,d_n|\bY,\bftheta) & = \prod_{i=1}^n p(\bu_i,d_i|\bY,\bftheta) = \prod_{i=1}^n p(\bu_i|d_i,\bY,\bftheta) \, p(d_i|\bY,\bftheta)\\
& = \prod_{i=1}^n \normal(\bu_i|\hat\bu_i,d_i\bG_i) \, \mathcal{IG}(d_i|\widetilde\alpha_i,\widetilde\beta_i), \label{eq:udpost}
\end{align}
where $\hat\bu_i = \bG_i \bX_i'\by_i$, $\bG_i = (\bX_i'\bX_i + \bV_i^{-1})^{-1}$, $\widetilde\alpha_i = \alpha_i + N/2$, and $\widetilde\beta_i = \beta_i + (\by_i'\by_i - \hat\bu_i' \bG_i^{-1} \hat\bu_i)/2 = \beta_i + (\by_i'(\bI_N+\bX_i \bV_i \bX_i')^{-1}\by_i)/2$.

Using \eqref{eq:udpost}, we can easily obtain samples or posterior summaries of the entries of $\bU$ and $\bD$ conditional on $\bftheta$. However, in many applications, primary interest will be in computing posterior summaries of $\bfSigma$ and other quantities. If $n$ is not too large ($n < 10^4$, say), we can simply compute $\bfSigma^{-1}$ (and hence $\bfSigma$) from $\bU$ and $\bD$. For large $n$, it is often not possible to even hold the entire dense matrix $\bfSigma$ in memory, but we can quickly compute useful summaries of it based on the sparse matrices $\bU$ and $\bD$ \citep[e.g.,][]{Katzfuss2018}. 
For example, a selected inversion algorithm can compute the variances $\bfSigma_{ii}$ and all entries $\bfSigma_{ij}$ for which $i \in g_m(j)$ or $j \in g_m(i)$. We can also compute the covariance matrix for any set of linear combinations $\bH\by^{(\ell)}$ as $\bH\bfSigma\bH'= \bA'\bA$, where $\bA=\bD^{1/2}\bU^{-1}\bH'$.
In many applications, including climate-model emulation, it is of interest to sample new spatial fields from the model, which we can do by sampling $\bz \sim \normal(\bfzero,\bI_n)$, and then setting $\by^\star = (\bU')^{-1}\bD^{1/2}\bz$; if $\bU$ and $\bD$ are sampled from their posterior distribution given $\bY$, then we have obtained a sample from the posterior predictive distribution $p(\by^\star|\bY)$.

\subsection{Parameterization of the prior distributions\label{sec:shrink}}

We now discuss parameterizing the NIG priors for $\bu_i$ and $d_i$ in \eqref{eq:nig} as a function of a small number of hyperparameters, $\bftheta = (\theta_1,\theta_2,\theta_3)'$, inspired by the behavior of Mat\'ern-type covariance functions. The parameter $\theta_1$ is related to the marginal variance, while $\theta_2$ and $\theta_3$ are related to the range and smoothness. In general, our prior parameterizations are motivated by interpreting $\bu_i$ and $d_i$ as the kriging weights and variance, respectively, for the spatial prediction problem implied by \eqref{eq:screening}, consisting of predicting $y_i^{(\ell)}$ from $\by_{g_m(i)}^{(\ell)}$; due to the maximin ordering, the locations of the variables in $\by_{g_m(i)}^{(\ell)}$ all have roughly similar distance to $\bs_i$ (see Figure \ref{fig:ordering}), and this distance decreases systematically with $i$.

\begin{figure}
    \centering
    \includegraphics[width=.78\textwidth]{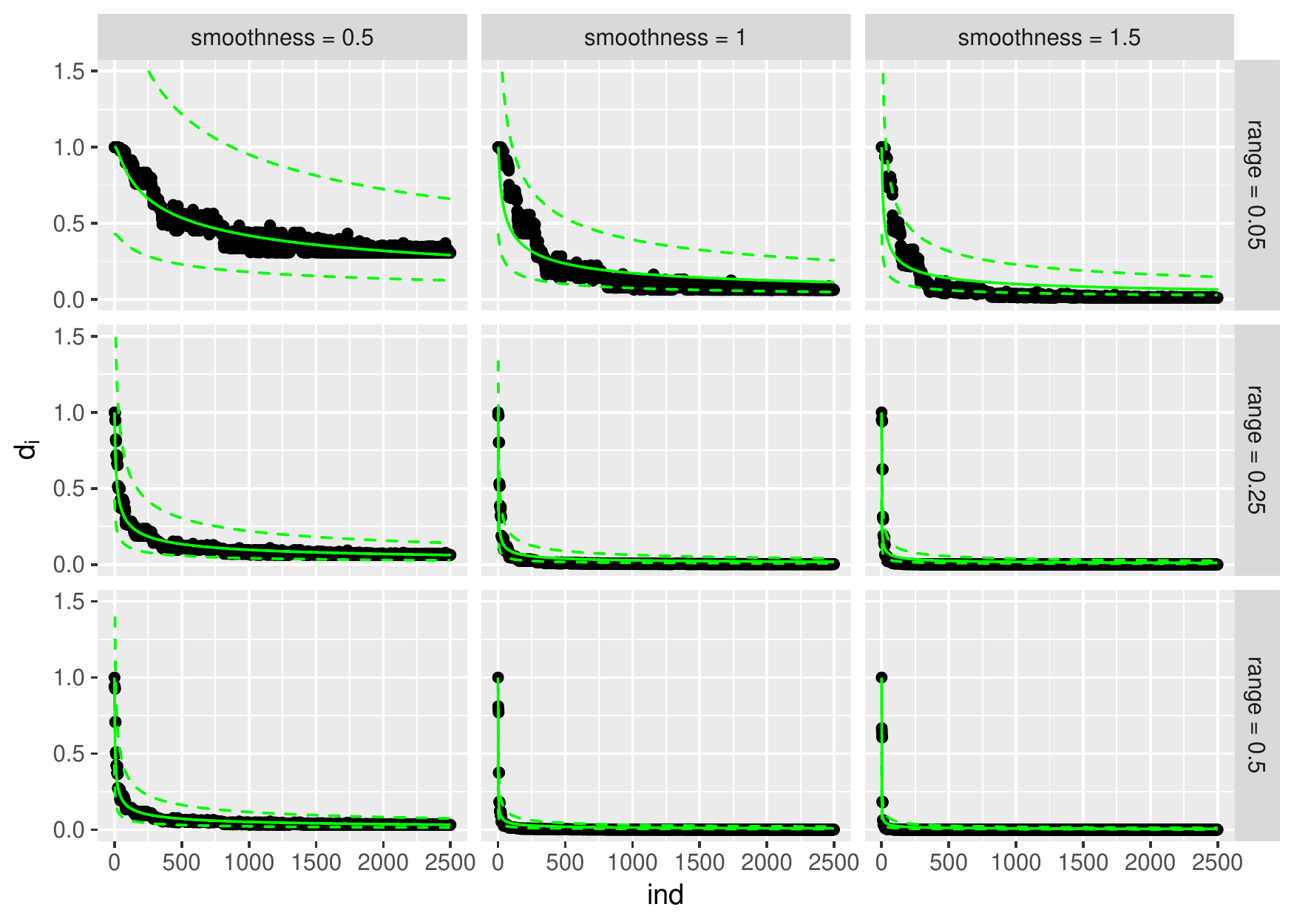}
    \caption{Illustration of the true entries $d_1,\ldots,d_n$ of $\bD$ as a function of location index $i$ for a Mat\'ern covariance function on a regular $n=50 \times 50 = 2{,}500$ grid on the unit square. The columns correspond to smoothness parameters, while the rows correspond to range parameters. The dashed lines are approximate 95\% pointwise intervals implied by our inverse-gamma prior, where $\theta_2$ was chosen for illustration using a least-squares fitting procedure (\texttt{nls} in \texttt{R}) assuming known $\theta_1=1$.}
    \label{fig:D}
\end{figure}

First, consider $d_i \sim \mathcal{IG}(\alpha_i,\beta_i)$ in \eqref{eq:nig}. For an exponential covariance with variance $\theta_1$ and range $2/\theta_2$, we have $\bfSigma_{i,j} = \theta_1 \exp(-\theta_2\|\bs_i-\bs_j\|/2)$; assuming $m=1$, we obtain
\begin{align}
d_i &= \var(y_i^{(\ell)}|\by_{g_m(i)}^{(\ell)}) 
= \theta_1 - {\textstyle\frac{(\theta_1 \exp(-\theta_2\|\bs_i-\bs_g\|/2))^2}{\theta_1}}
= \theta_1(1-e^{-\theta_2\|\bs_i-\bs_g\|}), \label{eq:dfun}
\end{align}
where $g = g_1(i)$, and the distance $\|\bs_i-\bs_g\|$ between location $\bs_i$ and its nearest previously ordered neighbor decreases roughly as $(i)^{-1/p}$ for a regular grid on a unit hypercube, $\domain = [0,1]^p$. (Throughout, $i$ is an index and not the imaginary number.) This motivates a prior for $d_i$ that shrinks toward $d_i \approx \theta_1(1-e^{-\theta_2(i)^{-1/p}})$. While \eqref{eq:dfun} only holds exactly for an exponential covariance with $m=1$, Figure \ref{fig:D} illustrates that this functional form approximately holds for Mat\'ern covariance functions in two dimensions with $m=n-1$ as well. 
Thus, we set the prior mean as $E(d_i|\bftheta) = \beta_i/(\alpha_i-1)= \theta_1 f_{\theta_2}(i)$, where $f_{\theta_2}(i) = 1-e^{-\theta_2(i)^{-1/p}}$. 
In Figure \ref{fig:D}, the empirically observed variance of the $d_i$ elements around the fit line decreases with $i$ as well, and so we set the prior standard deviation of $d_i$ to be half of the mean. Solving for $\alpha_i$ and $\beta_i$, we obtain $\alpha_i = 6$ and $\beta_i = 5\theta_1 f_{\theta_2}(i)$, because $Var(d_i | \bftheta) = \beta_i^2/((\alpha_i-1)^2(\alpha_i-2))$. 

\begin{figure}
    \centering
    \includegraphics[width=.78\textwidth]{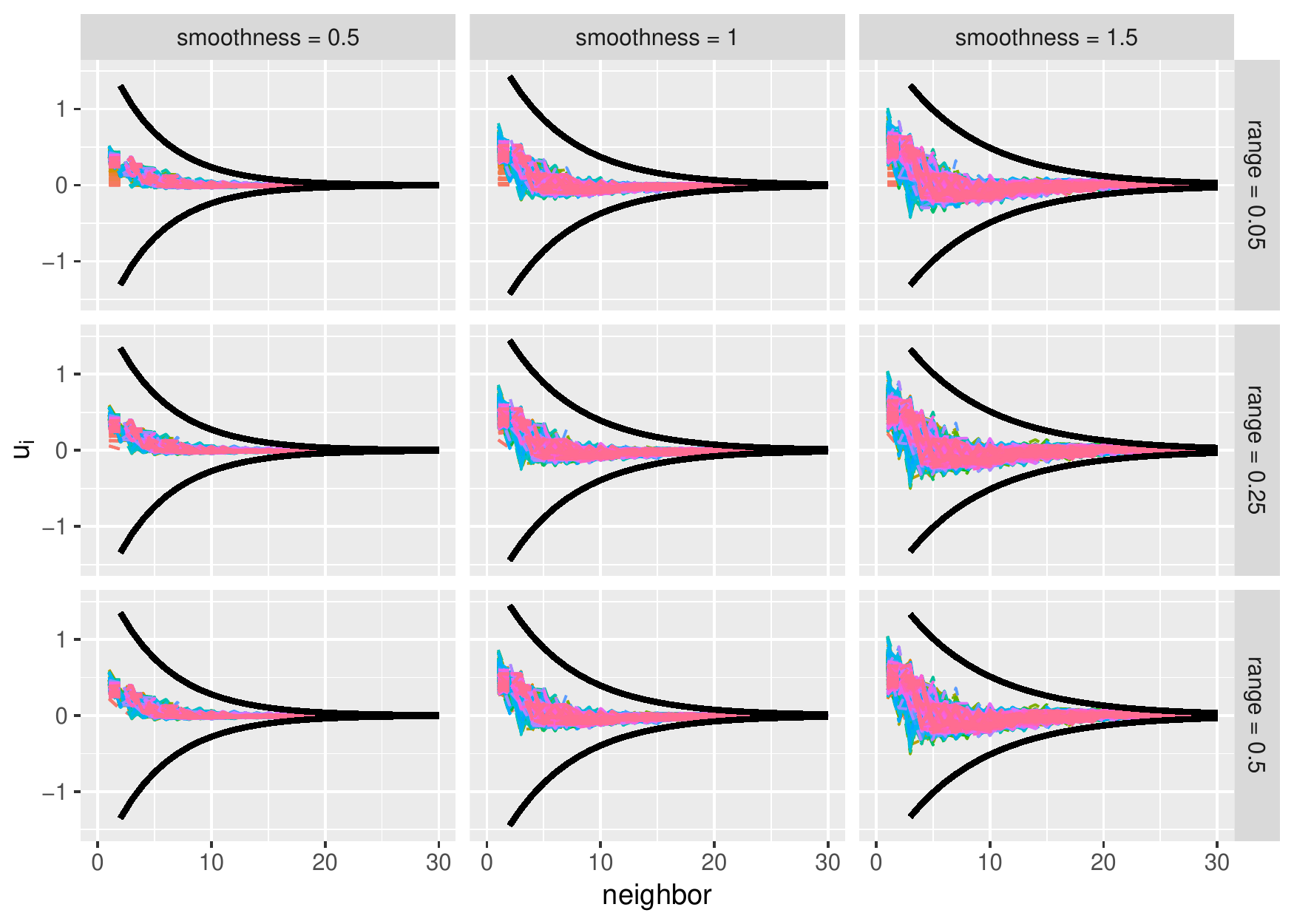}
    \caption{Illustration of the entries $(\bu_i)_j$ of $\bU$ as a function of neighbor number $j$ for the same setting as in Figure \ref{fig:D}. The dark lines correspond to approximate pointwise 95\% prior intervals ($\pm 2 \sqrt{\exp(-\theta_3 i)}$).} 
    \label{fig:U}
\end{figure}

Recent results based on elliptic boundary-value problems \citep[][Sect.~4.1.2]{Schafer2017} imply that the Cholesky entry $(\bu_i)_j$, corresponding to the $j$th nearest neighbor, decays exponentially as a function of $j$, for Mat\'ern covariance functions whose spectral densities are the reciprocal of a polynomial (ignoring edge effects). Thus, we assume $v_{ij} = \exp(-\theta_3 j) /( \theta_1 f_{\theta_2}(i))$ for $\bV_i = \diag(v_{i1},\ldots,v_{im})$ in $\bu_i | d_i,\bftheta \sim \normal(\bfzero,d_i\bV_i)$ in \eqref{eq:nig}. Note that we divide by $E(d_i|\bftheta)$ in $v_{ij}$, because the prior variance in $(\bu_i)_j|\bftheta \sim \normal(0,d_i v_{ij})$ is multiplied by $d_i$.
Figure \ref{fig:U} demonstrates this exponential decay as the neighbor number increases. 

Finally, consider the choice of conditioning-set size $m$. Simply setting $m$ to a fixed, reasonable value (e.g., $m \approx 10$, depending on computational constraints) works well in many settings, but the results can be highly inaccurate if $m$ is chosen too small, and the computational cost is unnecessarily high if $m$ is chosen too large. Hence, we prefer to allow the data to choose $m$ by tying $m$ to the prior decay of the elements of $\bU$; for all of our numerical experiments, we set $m$ as the largest $j$ such that $\exp(-\theta_3 j) > 0.001$, where $j$ denotes the neighbor number. This coincides with the amount of variation expected to be learnable from the data. 
Thus, entries of $\bU$ with sufficiently small prior variance as implied by a specific $\theta_3$ are set to zero, which ensures computational feasibility of our method.

\subsection{Inference on the hyperparameters $\bftheta$ \label{sec:hyperparams}}

The hyperparameters $\bftheta = (\theta_1,\theta_2,\theta_3)'$ determine $m$, $\bV_i$, $\alpha_i$, and $\beta_i$ as described in Section \ref{sec:shrink}. We now discuss how $\bftheta$ can be inferred based on the data $\bY$. All elements of $\bftheta$ are assumed to be positive due to the decay previously discussed, and so we perform all inference on the logarithmic scale.

The crucial ingredient for inference on $\bftheta$ is the marginal or integrated likelihood, which can be obtained by combining \eqref{eq:regression} and \eqref{eq:nig}, moving the product over locations outside of the integral over the entries of $\bU$ and $\bD$, and simplifying:
\begin{align}
    p(\bY|\bftheta) &= \prod_{i=1}^n \int_{d_i}\int_{\bu_i} \normal_N(\by_i|\bX_i\bu_i,d_i\bI_N) \normal(\bfzero,d_i\bV_i)  \mathcal{IG}(\alpha_i,\beta_i) d\bu_i dd_i \\
    &\propto \prod_{i=1}^n \, \Big( \, \frac{|\bG_i|^{1/2}}{|\bV_i|^{1/2}}  \, \frac{\beta_i^{\alpha_i}}{\widetilde\beta_i^{\widetilde\alpha_i}} \, \frac{\Gamma(\widetilde\alpha_i)}{\Gamma(\alpha_i)} \, \Big),    \label{eq:intlik}
\end{align}
where $\Gamma$ denotes the gamma function, the prior parameters $\alpha_i, \beta_i, \bV_i$ are given in \eqref{eq:nig}, and the posterior parameters $\widetilde\alpha_i, \widetilde\beta_i, \bG_i$ are given in \eqref{eq:udpost}. 

Based on this integrated likelihood, both empirical and fully Bayesian inference are straightforward.  Empirical Bayesian inference is based on a point estimate of $\bftheta$ obtained by numerically maximizing the log integrated likelihood. 
Fully Bayesian inference requires the specification of a hyperprior for $\bftheta$, which we simply assume to be flat (on the log scale). As a result, the posterior distribution $p(\bftheta|\bY) \propto p(\bY|\bftheta)$ is proportional to the integrated likelihood in \eqref{eq:intlik}. While this distribution cannot be obtained analytically, we can sample from the posterior using the Metropolis-Hastings (MH) algorithm. 
To avoid slow mixing due to large negative correlation between $\theta_1$ and $\theta_2$, we employ an adaptive MH algorithm that jointly proposes $\bftheta$ and learns its covariance matrix on-line; specifically, we use the implementation in \texttt{R} by \cite{scheidegger2012adaptmcmc}. 

\subsection{Computational complexity}

The cost for inference, including computing the posteriors in \eqref{eq:udpost}, sampling $\by^\star$, or evaluating the integrated likelihood in \eqref{eq:intlik}, is dominated by computing the $m\times m$ matrix $\bG_i$, which requires $\order(m^2N)$ time, and decomposing $\bG_i$, which requires $\order(m^3)$ time, for each $i=1,\ldots,n$. Hence, the time complexity is $\order(n(m^2N + m^3))$ for each unique value of $\bftheta$, where $m$ is often very small (e.g., $m \approx 10 $ in most of our numerical experiments). In addition, the most expensive computations can be carried out in parallel over $i=1,\ldots,n$.

For very small numbers of replicates, with $N <m$, we can use alternative expressions (see below \eqref{eq:udpost}) relying on computing and decomposing the $N \times N$ matrix $\bX_i \bV_i \bX_i' + \bI_N$ (instead of $\bG_i$), which requires $\order(mN^2 + N^3)=\order(mN^2)$ time.

The maximin ordering and large nearest-neighbor conditioning sets (with $m_{\max} = 50$, say) can be computed in quasilinear time in $n$ \citep{Schafer2017,Schafer2020}. For any $m \leq m_{\max}$ implied by a specific $\bftheta$, we can then simply select $g_m(i)$ as the first $m$ entries of $g_{m_{\max}}(i)$.

\subsection{Asymptotics}

Assume temporarily that \eqref{eq:datamodel} holds for some true $n \times n$ positive-definite covariance matrix $\bfSigma_0$, with fixed $n$ and $N \rightarrow \infty$. Then, the data model with the true $\bfSigma_0$ can be written in the regression form \eqref{eq:regression} with $m=n-1$. Under these assumptions, there are a fixed number (depending only on $n$, not on $N$) of variables in the regression models, and our prior distributions on the $\bu_i$, $d_i$, and $\bftheta$ place nonzero mass on the true model. Hence, using well-known asymptotic results based on the Bernstein--von Mises theorem \citep[e.g.,][]{VanderVaart2000}, the posterior distributions will be asymptotically normal and our posterior of $\bfSigma$ will contract around the true covariance $\bfSigma_0$ as the number of independent replicates $N$ approaches infinity. However, results of this nature are of limited use here, as we are most interested in the case $n \gg N$, which we will examine numerically in Sections \ref{sec:simulation} and \ref{sec:data}.

\subsection{Correlation-based ordering\label{sec:corr}}

For our methods, as discussed in Section \ref{sec:model}, we recommend a maximin ordering of the variables $y_1,\ldots,y_n$, and then selecting the conditioning sets $g_m(i)$ based on the $m$ nearest previously ordered variables, with $m$ determined by $\bftheta$ as described at the end of Section \ref{sec:shrink}. So far, these tasks were assumed to be based on the Euclidean distance of the corresponding locations $\bs_1,\ldots,\bs_n$ (see Figure \ref{fig:ordering}), which implies that our priors shrink toward isotropy (i.e., distributions for which dependence is only a function of distance). This shrinkage is not appropriate in some real-data applications. However, it is relatively straightforward to adapt our methods to processes (e.g., anisotropic or nonstationary) for which Euclidean distance is not meaningful. We merely require some prior guess of the correlation structure, based on expert knowledge, historical data, or (a regularized version of) the sample correlation of the data $\bY$; a simple choice used here is the element-wise product of the sample correlation and an (isotropic) exponential correlation with a large range parameter (e.g., half the maximum distance between any pair of locations in the dataset). Then, our procedures can be carried out as before, except that the ordering and nearest-neighbor conditioning is based on a correlation distance, defined as $(1 - |\text{correlation}|)^{1/2}$. This implicitly scales the space, so that the process is approximately isotropic in the transformed space. This approach can increase accuracy in the context of Vecchia approximations of parametric covariances (Kang and Katzfuss, in prep.); we propose it here for our nonparametric procedures. \citet[][Alg.~7]{Schafer2020} allows us to compute the correlation-based ordering and conditioning sets in quasilinear time in $n$.

\subsection{Noise or spatial trend\label{sec:noise}}

Our methodology described so far is most appropriate if the data are observed without any noise or nugget, meaning that realizations of the underlying spatial field are continuous over space; in this setting, approximations based on sparse inverse Cholesky factors of many popular covariance functions can be highly accurate \citep[e.g.,][]{Katzfuss2017a,Schafer2020}.

Now consider noisy observations $\bw^{(\ell)}| \by^{(\ell)} \stackrel{iid}{\sim} \normal_n(\by^{(\ell)},\tau^2\bI_n)$, $\ell=1,\ldots,N$, with $\by^{(\ell)}$ as in \eqref{eq:datamodel}. One option is to simply apply our methodology directly to the data $\bw^{(\ell)}$ as before; this will likely work well if the noise variance $\tau^2$ is small, but the conditional-independence assumption in \eqref{eq:screening} is less appropriate if $\tau^2$ is large \citep[e.g.,][]{Katzfuss2017a}, meaning that a much larger $m$ might be necessary. A larger $m$ results in higher computational cost and potentially less accuracy due to the higher number of Cholesky entries that must be estimated.

Hence, for large noise levels, we instead propose a Gibbs sampler that iterates between sampling $\by^{(\ell)}$ conditional on $\bw^{(\ell)}$ and $\bfSigma^{-1} = \bU \bD^{-1}\bU'$, and sampling $\bftheta$ and the entries of $\bU$ and $\bD$ conditional on the $\by^{(\ell)}$ as in Sections \ref{sec:predictive} and \ref{sec:hyperparams}. 
The former task can be accomplished without increasing the computational complexity for each Gibbs iteration, by exploiting the sparsity of the Cholesky factor $\bU \bD^{-1/2}$ of the prior precision, and approximating the Cholesky factor of the posterior precision using an
incomplete Cholesky factorization to avoid fill-in as described in \citet[][Sect.~4.1]{Schafer2020}.
(If $\tau^2$ is unknown, it is straightforward to sample from its full-conditional distribution as well.) 

A similar Gibbs-sampling strategy can be employed to make inference on a spatial trend. For example, if the observations $\bw^{(\ell)}$ are given by $\by^{(\ell)}$ plus a linear spatial trend with a Gaussian prior on the trend coefficients, the coefficients can be sampled in closed form conditional on $\bfSigma$, and all other unknown quantities can be sampled given the trend coefficients as before based on $\by^{(\ell)}$ obtained by subtracting the trend from $\bw^{(\ell)}$.

\section{Simulation study \label{sec:simulation}}

We compared the following methods:
\begin{description}[itemsep=1pt,topsep=2pt,parsep=1pt]
\item[SCOV:] Basic sample covariance
\item[OURS:] Our method described in the previous sections
\item[MLE:] Estimate based on the MLEs of $\bu_i$ and $d_i$ for the regressions in \eqref{eq:regression} (i.e., no prior shrinkage), with $m = \min(m_{\text{OURS}},N-1)$, with $m_{\text{OURS}}$ implied by OURS $\bftheta$ estimate
\item[LASSO:] Lasso for each regression in \eqref{eq:regression}, with all possible previous points included as possible predictors (i.e., $m=n-1$)
\item[SLASSO:] Spatial LASSO with penalty scaled by the spatial distance to favor inclusion of nearer points as predictors, intended to be similar to \citet{zhu2009estimating}
\end{description}
The spatial domain for all comparisons was the unit square.

\subsection{Uncertainty quantification\label{sec:uq}}

First, we fit a fully Bayesian version of OURS to simulated data, to demonstrate the uncertainty quantification in the covariance estimation. Specifically, we considered $N=20$ realizations of a Gaussian process with Mat\'ern covariance function with variance 3, smoothness 1, and range parameter 0.25, at $n=900$ randomly sampled locations. We obtained 50,000 samples of $\bftheta$ using an adaptive MCMC \citep{scheidegger2012adaptmcmc}. The trace plots showed good mixing and convergence, and the individual effective sample sizes for the three parameters were all larger than 1,000. After conservatively discarding the first half of the samples for burn-in and thinning by a factor of 50, a covariance matrix was calculated from a sample from \eqref{eq:udpost} for each $\bftheta$ draw.

\begin{figure}%
    \centering
	\begin{subfigure}{.48\textwidth}
	\centering
\includegraphics[width=0.99\linewidth]{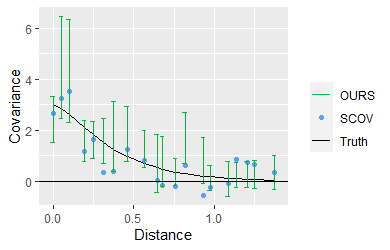}
	\caption{}
	\label{fig:CIsample}
	\end{subfigure}%
\hfill
	\begin{subfigure}{.48\textwidth}
	\centering
\includegraphics[width=0.99\linewidth]{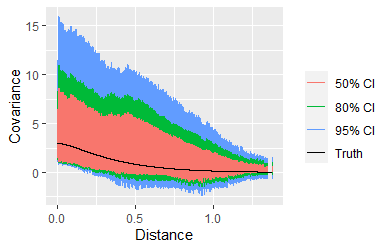}
	\caption{}
	\label{fig:CIall}	
	\end{subfigure}%
    \caption{Based on $N=20$ draws from a Gaussian process with Mat\'ern covariance at $n=900$ locations (see Section \ref{sec:uq}): (a) Sample estimates (SCOV) and posterior 80\% credible intervals using our fully Bayesian method (OURS) for 20 entries of the covariance matrix. (b) 50\%, 80\%, and 95\% credible intervals using OURS for one randomly sampled entry of the covariance matrix corresponding to each unique distance.}%
    \label{fig:CI}%
\end{figure}

Figure \ref{fig:CIsample} shows the resulting 80\% posterior credible intervals (CIs) along with the SCOV estimates for 20 randomly sampled matrix entries $\bfSigma_{ij}$ as a function of $\|\bs_i - \bs_j\|$, the distance between the corresponding spatial locations. Most of the OURS CIs contained the true value and tracked the decay of the covariance as a function of distance. This is also the general trend for CIs at all distances shown in Figure \ref{fig:CIall}.

\subsection{Comparison to LASSO for small $n$\label{sec:compsmalln}}

We compared estimation accuracy using the Kullback-Leibler (KL) divergence between the estimated distribution $\normal_n(\bfzero,\hat\bfSigma)$ and the true distribution $\normal_n(\bfzero,\bfSigma)$:
$$
\KL(\hat{\bfSigma} \| \bfSigma) = \tr(\hat{\bfSigma} \bfSigma^{-1}) - \log |\hat{\bfSigma}\bfSigma^{-1}| - n,
$$
where $\tr(\cdot)$ denotes the trace and $| \cdot |$ denotes the determinant. This exclusive KL divergence does not require inverting the estimate $\hat{\bfSigma}$ and thus avoids issues with SCOV for $N < n$. To obtain a point estimate for OURS, we computed $\hat{\bfSigma} = (\hat\bU^{-1})' \hat\bD\hat\bU^{-1}$, where $\hat\bU$ and $\hat\bD$ were the maximum a posteriori (MAP) estimates from \eqref{eq:udpost}, based on the value of $\bftheta$ that maximized the integrated likelihood \eqref{eq:intlik}.


\begin{figure}
    \centering
    \includegraphics[width=0.65\linewidth]{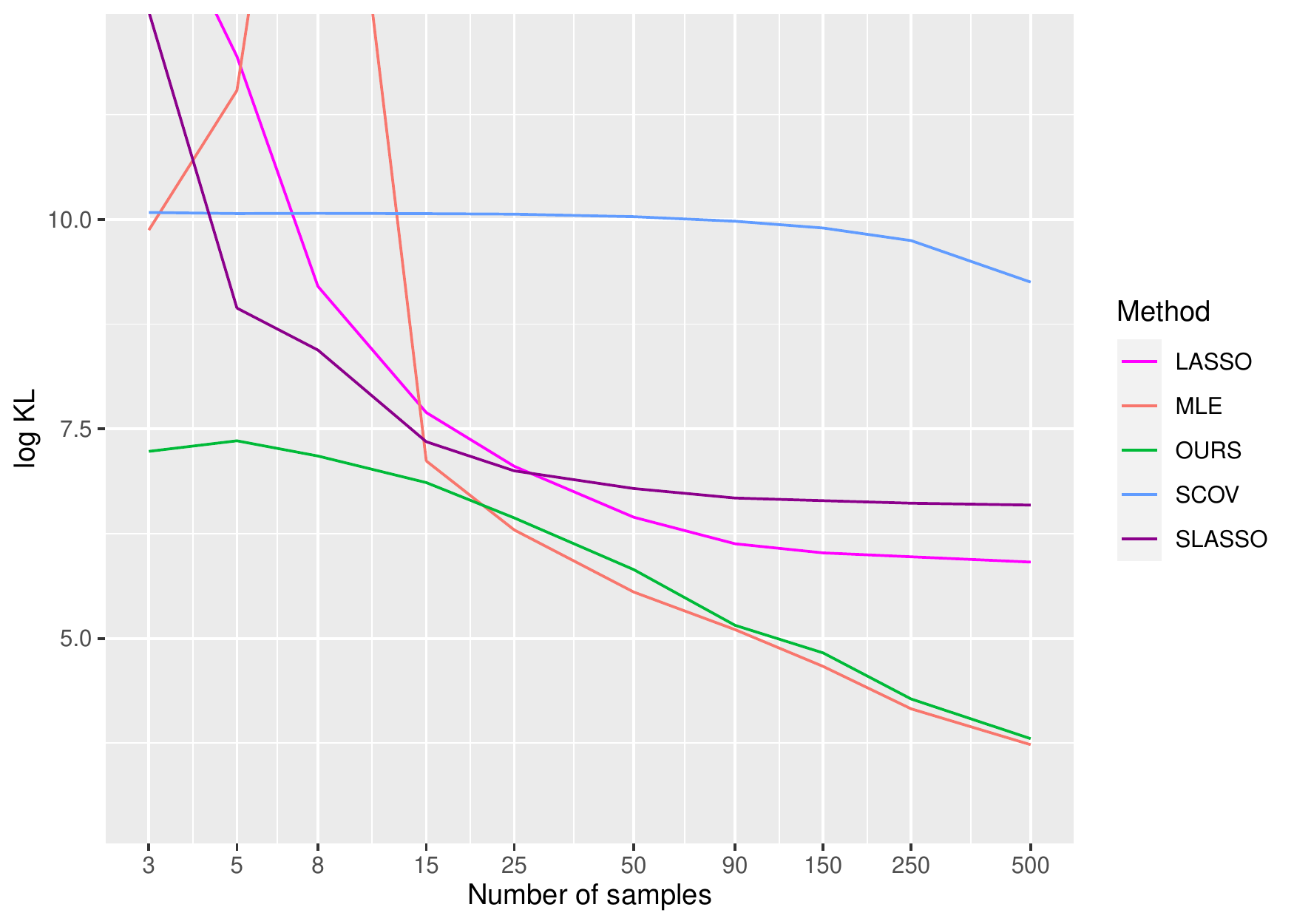}
    \caption{For the comparison in Section \ref{sec:compsmalln}, KL divergence (on a log scale) for different covariance estimation methods for varying numbers $N$ of samples from a Mat\'ern covariance at $n=900$ locations}
    \label{fig:glm}
\end{figure}

Figure \ref{fig:glm} shows the results, using the same set-up with $n=900$ as in Section \ref{sec:uq}, for various numbers of replicates $N$. MLE was similarly accurate as OURS for large $N$, as expected, but it performed worse for small $N$ due to the lack of prior shrinkage. Similarly, the inclusion of spatial information in SLASSO resulted in higher accuracy than LASSO for small $N$. LASSO and SLASSO were not competitive with OURS and MLE, despite increased flexibility in selecting predictors (i.e., conditioning sets) in the regressions \eqref{eq:regression}, and despite much higher computational cost due to calculations involving all $O(n)$ possible predictors. Hence, we did not consider (S)LASSO further.

\subsection{Comparison for larger $n$\label{sec:complargen}}

\begin{figure}
    \centering
    \includegraphics[width=0.95\linewidth]{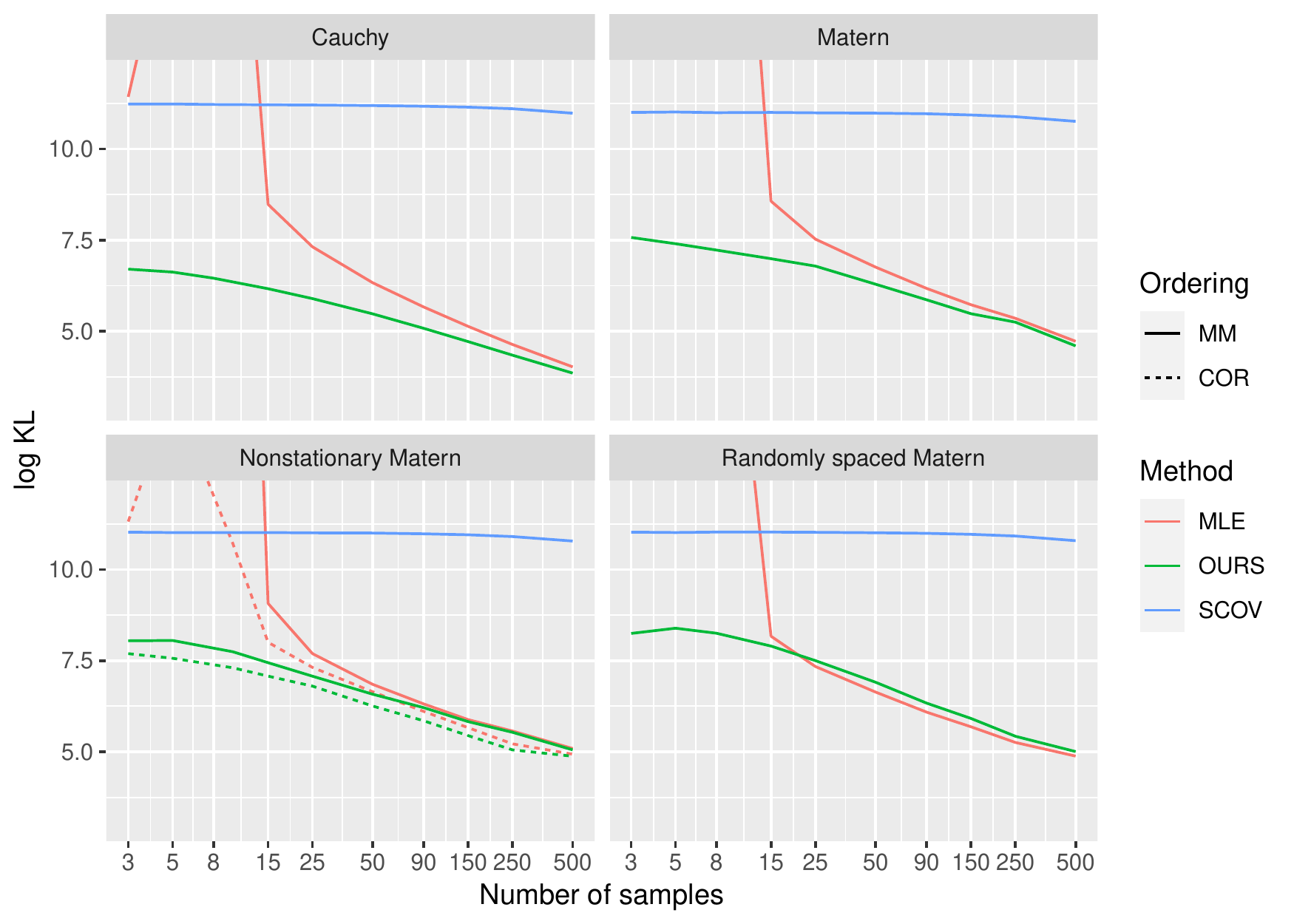}
    \caption{Comparison of KL divergence (on a log scale) for four different settings with $n=2{,}500$ described in Section \ref{sec:complargen}. Correlation-based ordering was only used for the nonstationary setting.} 
    \label{fig:sims}
\end{figure}

Figure \ref{fig:sims} shows further comparisons with $n=2{,}500$ spatial locations using the KL divergence in four different settings (counter-clockwise from top right), all with a marginal variance of 5: Mat\'ern with smoothness 1 and range parameter 0.5 on a regular $50 \times 50$ spatial grid (corresponding to the middle panel in the bottom row of Figures \ref{fig:D} and \ref{fig:U});
a Cauchy covariance with range 0.25 and memory parameters 1 and 0.5 on a regular $50 \times 50$ grid; 
Mat\'ern covariance with varying anisotropy \citep{Paciorek2006}, for which the range parameter is constant at 0.05 in the $x$ direction but varies as $0.05 + 0.45 \, s_y$ (as a function of the $y$-coordinate $s_y$) in the $y$ direction, on a regular $50 \times 50$ grid;
Mat\'ern with smoothness 1 and range 0.25 at $n=2{,}500$ randomly spaced locations sampled uniformly.

For all scenarios, MLE was roughly as accurate as OURS for very large $N$, but performed poorly for small $N$, indicating that the added shrinkage from our prior improved the accuracy. OURS strongly outperformed SCOV in all settings.
For the nonstationary covariance, we also considered the correlation-based ordering (COR) described in Section \ref{sec:corr}. While we used the true correlation for the comparison here, the element-wise product of the sample covariance and an exponential correlation proposed in Section \ref{sec:corr} resulted in comparable accuracy (not shown).
As expected, OURS-COR performed better than OURS-MM in this nonstationary setting. 
We also conducted some experiments (not shown) using a natural ordering by one of the spatial coordinates, which performed comparably to maximin ordering for isotropic covariances on a regular grid, but was much less accurate for randomly sampled locations.

Overall, our method performed well across all simulations, even though our prior distributions were motivated by isotropic Mat\'ern-like covariances. In addition, the computational burden for OURS was relatively low, with the estimated $m$ often around ten and always below 30.
While we only considered moderate $n$ here in order to be able to carry out many comparisons using the KL divergence, it is also possible to run our method on much larger datasets. For example, using a C++ implementation, evaluating the integrated likelihood \eqref{eq:intlik} once only took about 6 seconds on a 4-core laptop (Intel i7-7560U) for $n=250{,}000$, $m=10$, $N=50$.

\section{Climate-model emulation\label{sec:data}}

We analyzed climate-model output from the Community Earth System Model (CESM) Large Ensemble Project \citep{Kay2015}. 
Specifically, we considered daily mean surface temperature (in Kelvin) on July 1 in $98$ consecutive years starting in the year 402, on a roughly $1^\circ$ longitude-latitude grid of size $n = 81 \times 96 = 7{,}776$ containing much of the Americas (see Figure \ref{fig:tempens}). The chosen region features ocean, land, islands, and mountain ranges, leading to a complicated, nonstationary dependence structure. 
The data $\bY$ were defined as the temperature anomalies obtained by standardizing the climate-model output at each grid point to unit mean and variance. 
We found no evidence of temporal correlation in the data, and so the assumption of independent replicates in \eqref{eq:datamodel} was at least approximately satisfied.

First, we compared three covariance estimates: an exponential covariance with a range parameter estimated from the data (EXP); a tapered sample covariance given by the element-wise product of the sample covariance and an exponential correlation with a range of 6,000 km, with a small added nugget with variance $10^{-5}$ for numerical stability (SCOVT); and the MAP estimate (as in Section \ref{sec:compsmalln}) using our method with correlation ordering (Section \ref{sec:corr}) based on the SCOVT matrix (OURS). Of the 98 replicates (i.e., years), we randomly selected and withheld 18 as test data, and fit the models on subsets of various sizes $N$ between 6 and 80. As the true distribution was unknown, it was not possible to compute the KL divergence. Instead, we used the strictly proper log score \citep[e.g.][]{Gneiting2014} given by the average negative log posterior predictive density of the test data based on \eqref{eq:datamodel}, with $\bfSigma$ replaced by the corresponding estimate for each of the three estimates.

\begin{figure}
    \centering
    \includegraphics[width=.7\linewidth]{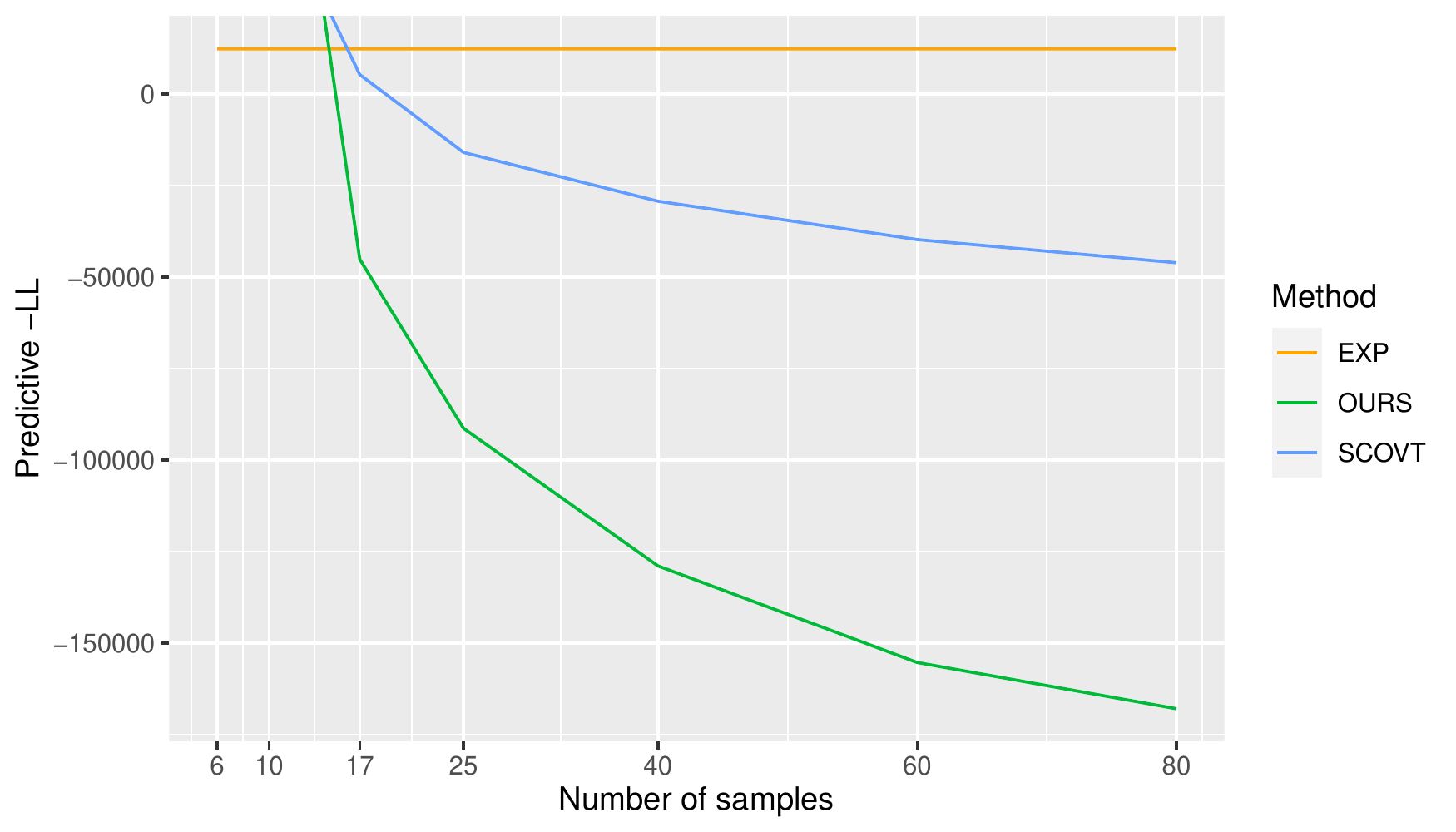}
    \caption{Comparison using the log score (lower is better) of methods fitted on climate-model temperature anomalies with varying numbers of replicates $N$ (see Section \ref{sec:data})}
    \label{fig:realdata}
\end{figure}

Figure \ref{fig:realdata} shows the resulting scores, averaged over five random training/test splits. OURS was more accurate than SCOVT for all values of $N$, and more accurate than EXP for all $N \geq 10$. We also tried OURS with Euclidean (instead of correlation-based) ordering, which resulted in similar scores for large $N$ but required almost twice the $N=17$ replicates to surpass EXP (not shown). While it may be possible to find other (e.g., parametric nonstationary) methods that can result in even lower scores than OURS for this dataset, such methods would likely require substantial amounts of manual tuning (e.g., specifying the parametric form of the nonstationarity). 

\begin{figure}
    \centering
    \includegraphics[width=.9\linewidth]{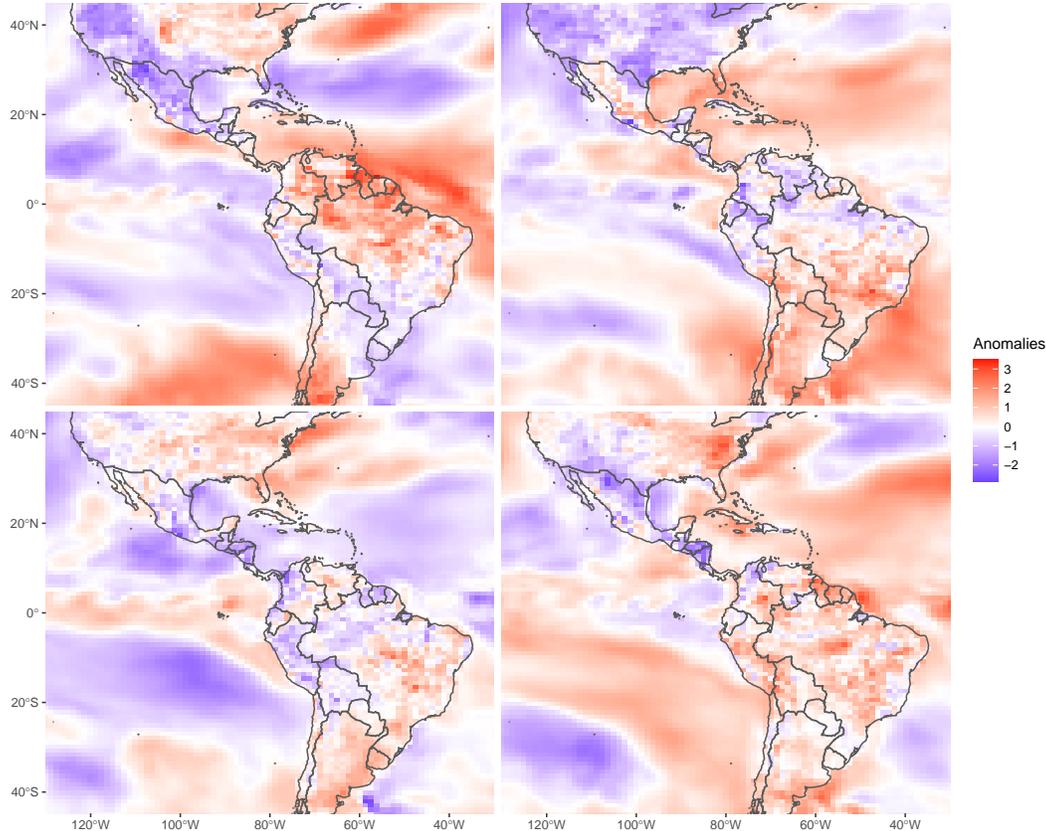}
    \caption{Four temperature-anomaly fields (in Kelvin) sampled from the posterior predictive distribution using our fully Bayesian method, computed as described in Section \ref{sec:data} based on climate-model output as in Figure \ref{fig:tempens}}
    \label{fig:tempsamp}
\end{figure}

We created a stochastic simulator emulating the climate model, by fitting a fully Bayesian version of OURS to the full dataset with $N=98$ and sampling from the posterior predictive distribution $p(\by^\star|\bY)$ as described at the end of Section \ref{sec:predictive}. Four such samples are shown in Figure \ref{fig:tempsamp}; they look qualitatively similar to the actual samples from the climate model in Figure \ref{fig:tempens}, including reproducing features corresponding to land/ocean effects despite using no explicit information on land boundaries. These results were based on 50,000 Metropolis-Hastings (MH) samples of $\bftheta$ (after a burn-in of 50,000) with trace plots showing good mixing and effective sample sizes all larger than 1,000; the samples were then thinned by a factor of 50. It took about 200 minutes to train the emulator, and it took 2.5 seconds to obtain a sample $\by^\star$ for a given value of $\bftheta$, on a 4-core laptop (Intel i7-7560U) without parallelization. 

\section{Conclusions \label{sec:conclusions}}

We have developed a scalable, flexible Bayesian model for spatial covariance estimation and emulation. We regularize our method by taking advantage of a form of ordered conditional independence often assumed for spatial data. This motivates the assumption of sparsity in the Cholesky of the precision matrix, which greatly improves scalability and reduces the number of unknown parameters from quadratic to near-linear in the number of spatial locations. We describe three hyperparameters related to the marginal variance and the decay of Cholesky entries; these hyperparameters can be quickly optimized or sampled, resulting in an automatic data-based selection of the sparsity structure. Hence, our method requires no manual tuning or cross-validation. While our approach was motivated by the behavior of isotropic covariances on regular grids, our numerical comparisons demonstrated its generality with more complex covariances and irregularly spaced locations. We also applied the method to climate-model emulation, where it captured the nonstationary behavior better than standard methods. 
Template \texttt{R} code for our implementation is provided with this article.

There are several interesting extensions for our spatial covariance estimation procedure.
Our method can be extended to handle missing values by imputation using a Gibbs sampler similar to the ones described in Section \ref{sec:noise}; however, if the number of observations at a particular location is very small or even zero, the posterior distribution at that location will be very vague and thus generally not particularly useful, unless some additional assumptions about the covariance between the unobserved and observed locations are made.
For example, more explicit shrinkage toward a specific parametric covariance could be achieved by setting the prior mean of the nonzero entries of $\bU$ and $\bD$ in \eqref{eq:nig} to the values implied by a parametric Vecchia approximation \citep[e.g.,][Sec.~4.1]{Katzfuss2017a}.
Another potential extension is to estimate the covariance as a function of external variables by including them as additional covariates in the regressions in \eqref{eq:regression}; for instance, for climate-model emulation, the covariance could depend on season, year, elevation, or land versus ocean.
Finally, our approach could be extended to data assimilation, by using it to infer the forecast covariance matrices in an ensemble Kalman filter.

\footnotesize
\appendix
\section*{Acknowledgments}

Katzfuss's research was partially supported by National Science Foundation (NSF) Grants DMS--1654083, DMS--1953005, and CCF--1934904.
We would like to thank Mohsen Pourahmadi, Florian Sch\"afer, Will Boyles, and Joe Guinness for helpful comments and suggestions.

\footnotesize
\bibliographystyle{apalike}
\bibliography{mendeley,additionalrefs}

\end{document}